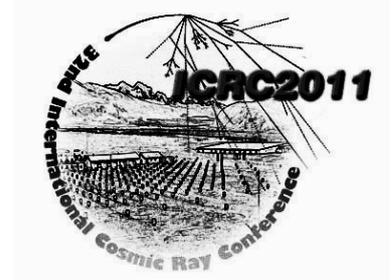

# Energy spectrum and mass composition of primary cosmic radiation in the region above the knee from the GAMMA experiment

R.M.Martirosov[1], A.P.Garyaka[1], H.S.Vardanyan[1], A.D.Erlykin[2], N.M.Nikolskaya[2], Y.A.Gallant[3], L.W.Jones[4], H.A.Babayan[5]

[1]A. Alikhanyan National Science Laboratory, Yerevan, Armenia
[2]P N Lebedev Physical Institute, Moscow, Russia
[3]LUMP, Université Montpellier II, Montpellier, France
[4]Department of Physics, University of Michigan, USA
[5]State Engineering University of Armenia, Yerevan, Armenia
romenmartirosov@rambler.ru

**Abstract.** The energy spectrum of the primary cosmic radiation in the energy range 1 – 100 PeV and the extensive air shower (EAS) characteristics obtained on the basis of the expanded data bank of the GAMMA experiment (Mt. Aragats, Armenia) are presented. With increased statistics we confirm our previous results on the energy spectrum. The spectral index above the knee is about -3.1, but at energies beyond 20 PeV a flattening of the spectrum is observed. The existence of the 'bump' at about 70 PeV is confirmed with a significance of more than 4σ. In the energy range of 10 – 100 PeV the shower age becomes energy independent and we observe a direct proportionality of the EAS size to the primary energy. This suggests an approximately constant depth of the EAS maximum in this energy range. This is evidence in favour of an increasing average mass of primary particles at energies above 20 PeV. The additional source scenario, which is a possible explanation of the 'bump' in the spectrum, also leads to the conclusion of increasing mass of the primary cosmic rays. A comparison with the data of other experiments is presented.

## 1. Introduction

The main goal of the study of the energy spectrum and mass composition of primary cosmic rays is an understanding of the mechanism of particle acceleration and propagation in the space. To understand the origin of the knee at 3-4 PeV is an important part of this goal. It is predicted that if cosmic rays are confined in supernova remnants and accelerated at strong shocks through the Fermi process, cosmic rays are enriched by heavy nuclei (like iron) as the primary energies increase. Ground-based installations that investigate electromagnetic and muon components of EAS realize this task over a wide energy range. The models [1-4] for the origin of the knee predict changes in the behavior of the all-particle spectrum and constituent component spectra. We need high accuracy experimental data on the EAS components to distinguish the correct model between hosts of them. It is commonly believed that the all-particle flux changes smoothly without any prominent structures. Some of the models predict the appearance of structures in the energy spectra beyond the knee. There are experimental indications that such structures exist [5]. The indirect character of the primary energy and mass determination demands an increase in the accuracy of the data treatment methods along with the increasing of the experimental data statistics. We present in this paper the all-particle primary energy spectrum and some EAS characteristic correlations on the basis of extended data from the GAMMA experiment. Special attention was paid for the checking of a possible influence of detector saturation on the derived energy spectrum.

## 2. GAMMA experiment

GAMMA is a ground-based EAS array to measure the muon and the electro-magnetic components of the EAS [6,7]. It consists of an array of 33 surface detection stations and underground muon detectors, located at the south side of Mount Aragats in Armenia. The layout of the array is shown in figure 1. Each station contains 3 plastic scintillation detectors with dimensions of 1x1x0.05 $m^3$. Each of the central nine stations contains an additional (the 4th) small scintillator with dimensions of 0.3x0.3x0.05 $m^3$ for high particle density ($>>10^3$ $m^{-2}$) measurements. In addition recently sepa-rate 8 detectors were placed to make central part denser. The muon carpet is composed of 150 scintillation detectors which are compactly arranged in the underground hall under the 2.3 kg/cm$^2$ of rock and concrete. The layout of the carpet is shown in figure 2. The scintillator dimensions, casings and photomultipliers are the same as in the surface detectors. The arrangement of the muon detectors gives the possibility of determining the muon lateral distribution up to 60m from the EAS core at $E_\mu > 5$GeV.

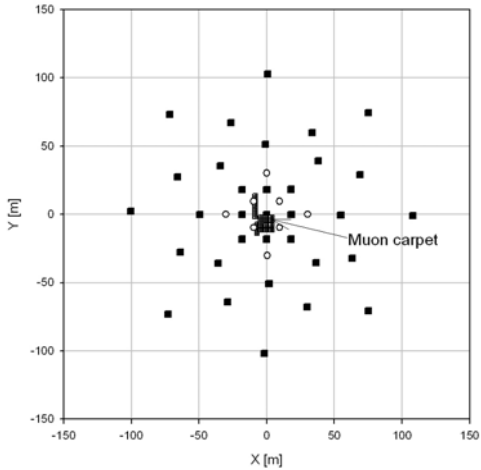

Figure 1. The layout of the GAMMA array. Black squares – detection stations. White circles – single detectors

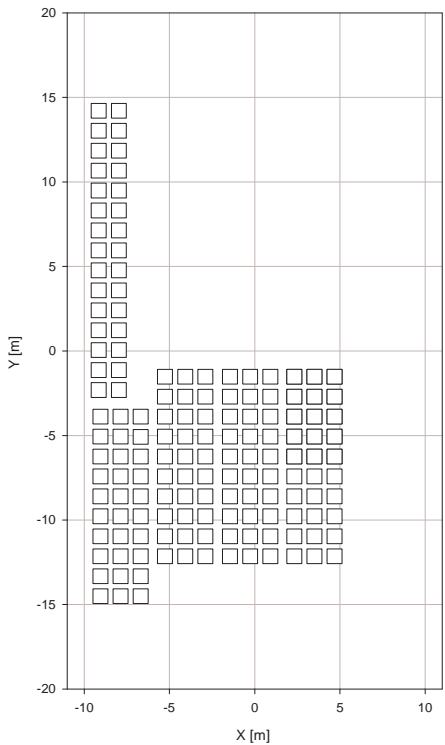

Figure 2. The layout of the muon carpet.

The reconstruction of the EAS size ($N_{ch}$), shower age ($s$) and core coordinates ($x_0, y_0$) is performed based on the Nishimura-Kamata-Greisen (NKG) approximation to the measured charged particle densities. Minimization was used to estimate $x_0, y_0$ and a maximum likelihood method to estimate $N_{ch}$, taking into account the measurement errors. The arrival direction of EAS was determined by timing measurements in surface detectors.

## 3. Results

Applying an event-by-event 7 parametric energy evaluation, the all-particle energy spectrum in the knee region is obtained on the basis of the data set obtained using the GAMMA EAS array data for the years 2003-2009, and a simulated EAS database obtained using the CORSIKA(NKG) EAS simulation code [8] with the SIBYLL [9] interaction model. The formula for the energy estimator is given in [7] together with the errors evaluation method description.

In the figure 3 we present the primary all-particle energy spectrum for the years 2003-2009 together with previously published [7]. In figure 4 the dependence of the average value of $s$ on $E_0$ is given. We compute total errors according to the method described in [7]. Core position selection criteria r < 20m $\theta < 21^0$ is used to decrease the lower limit of the spectrum. Two spectra are in accordance with small difference. The main features are the same. After the *knee* the spectrum has a power index -3.1 but at $E_0 > 10$ PeV it cannot be approximated by simple function. It deviates from a power law and goes higher to the bump structure at about 70 PeV where the deviation is more than 4σ. Also the *s* dependence on $E_0$ becomes roughly constant in the same region.

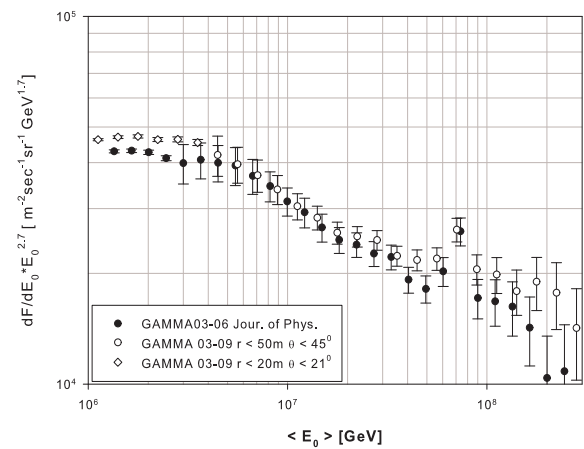

Figure 3. All-particle primary energy spectrum.

It was shown in [6,7] that rigidity-dependent primary energy spectra cannot describe such behavior and the presence of additional component in the primary flux of nuclei is necessary. The hypothesis of two-component origin of cosmic ray flux was used to ex-plain the fine structure of the spectrum. It took into account the contribution of pulsars in the Galactic cosmic ray flux. We carried out the test of this hypothesis [7].

The parameter $s$ is a lateral age parameter that is derived from our data. It is directly connected with the longitudinal age and at the same time the age is dependent on the distance from the depth of shower maximum to the observation level.

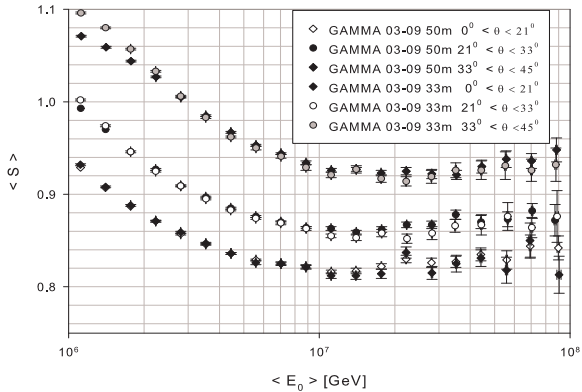

Figure 4. Age parameter dependence on $E_0$

Constancy of the age means the constancy of the depth of maximum (approximately). This is possible for a complex and changing primary composition if the flux became heavier with energy. The depth of maximum grows with energy as $\sim \ln E_0$ and (from the other side) decreases as the mean mass increases. In the superposition model approximation, it changes as $\sim <\ln A>$.

These two factors must compensate each other to result in the constancy of the age parameter. In some experiments such a rise in the $<\ln A>$ was observed [see review 10]. This consideration is rather qualitative. But though it predicted a rapid growth of $<\ln A>$ with energy.

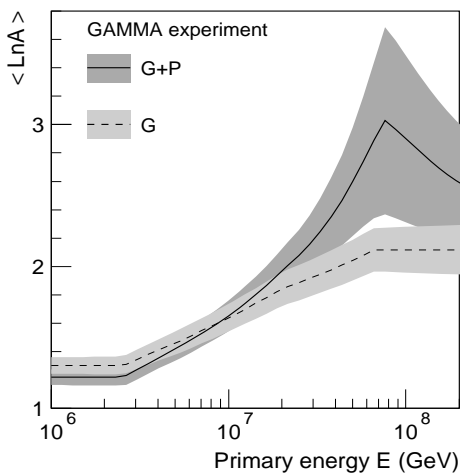

Figure 5. Average logarithm of primary mass number as an energy function. 2-component model prediction [7].

In the figure 5 an average logarithm of primary mass number as a function of energy is shown in the case of the hypothesis of two-component origin of the cosmic ray flux.

Such growth of the $<\ln A>$ is reflected on another correlation distribution, the $N_e$ dependence on $E_0$. It is shown in the figure 6. Although $N_e$ is used in the determination of $E_0$, it is only one among 7 other parameters.

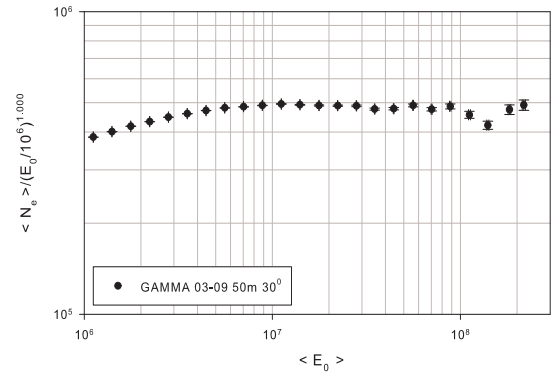

Figure 6. $N_e$ dependence on $E_0$

As a rule in the case of constant mass composition this dependence is $N_e \sim E_0^{1+\varepsilon}$ in the range after the shower maximum. Just such behavior is seen before the knee where $\varepsilon$ is about 0.13. After the knee $\varepsilon$ decreases and becomes about 0 at $E_0 > 10$ PeV. Such behavior is consistent with the approximate constancy of the depth of shower maximum.

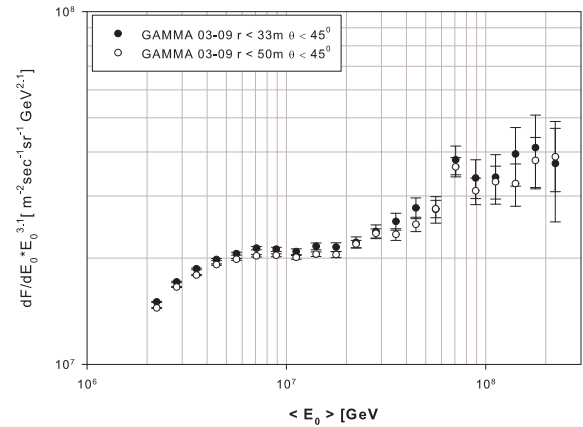

Figure 7. All-particle primary energy spectrum multiplied by $E_0^{3.1}$ in two ranges by distance from array centre.

We carried out a test to check the possible influence of the saturation of our big (1 sq.m) scintillation detectors at high charged particle density. In many detectors the saturation is reached at 2000 part/m². It means that we can not correctly determine the particle density near the core for event with $E_0 > 10$ PeV. It is possible only for showers with axes in the circle of radius 33 m where we have small detectors nearby EAS axes. In figure 7 we present energy spectrum obtained in two circles with radii 33 and 50 m. It is multiplied by $E_0^{3.1}$ to demonstrate the fine structure of the spectrum more clearly. It is seen that both spectra have the same behavior and bump structure. Also the average $s$ dependence on $E_0$ coincides in two ranges as it seen in the figure 4. It means that the saturation effect is not strong in our data.

In the same energy range there are preliminary results of TUNKA [11] and KASKADE-Grande [12] on the primary energy spectrum above the knee. Both data show deviation from the power law with index -3.1. In figure 8 we present our spectrum with the preliminary data of the TUNKA experiment. It is significant that data obtained with different methods are so close. In the same energy region the KASCADE-Grande resulting energy spectrum shows statistically significant flattening that can be identified with a concavity.

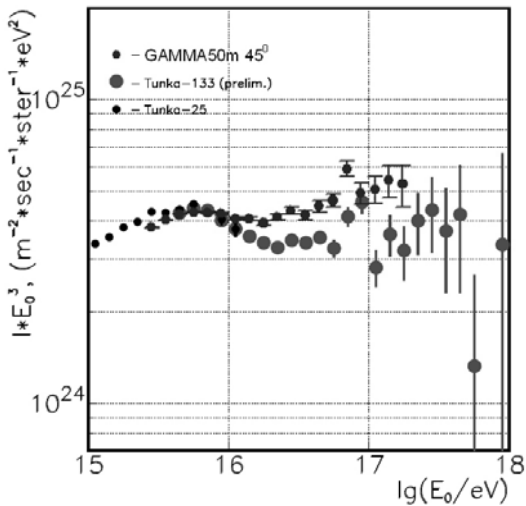

Figure 8. Comparison of GAMMA and Tunka [11] preliminary results.

## 4. CONCLUSIONS

The all particle primary energy spectrum was derived on the basis of GAMMA data for 2003-2009 years. We confirm our result [7]. At the same time in the energy range of 10-100 PeV the spectrum shows statistically significant feature of deviation (flattening) from the power law with index of -3.1. The observed bump can be described by a two-component model of the primary cosmic ray origin.

The constancy of the EAS age parameter s and the proportionality of the EAS size $N_e$ to the primary energy $E_0$ suggest the growth of the mean logarithm of the mass <lnA> of primary cosmic rays above the knee. Our results are in qualitative agreement with the data of Tunka experiment.

**Acknowledgments**

We are grateful to all our colleagues at the Moscow P N Lebedev Physical Institute and the A. Alikhanyan National Laboratory who took part in the development and exploitation of the GAMMA array. This work has been partly supported by the research grant no. 090 from the Armenian government and the 'Hayastan' All-Armenian Fund.